\documentclass[12pt,a4paper]{iopart}

\usepackage{iopams,setstack,amsthm}

\eqnobysec

\newcommand{\Ltwo}{{\rm L}^2}
\newcommand{\PT}{{\rm P}\Gamma_\theta}
\newcommand{\PGT}{{\rm P}\Gamma(2)}
\newcommand{\HN}{\mathcal{H}_N}
\newcommand{\SLTZ}{{\rm SL}(2,\mathbb{Z})}
\newcommand{\notequiv}{\not \hspace{.1mm} \equiv}
\newcommand{\Op}{\mathop{\mathrm{Op}_N}\nolimits}
\newcommand{\sign}{\mathop{\mathrm{sign}}\nolimits}
\newcommand{\abs}[1]{\left | #1 \right |}
\newcommand\Jsym[2]%
{\Biggl(\begin{array}{c} #1 \\ \hline #2 \end{array}\Biggr)}
\newcommand{\invint}[2]{\left( #1 \backslash #2 \right)}

\newcommand{\msc}[1]{\vspace{10pt}
     \begin{indented}
     \item[]\rm Mathematics Subject Classification: #1\par
     \end{indented}}

\newcommand{\mydate}[1]{\vspace{5pt}
        \begin{indented}
        \item[]\rm #1\par
        \end{indented}}

\newtheorem{theorem}{Theorem}[section]
\newtheorem{lemma}[theorem]{Lemma}
\newtheorem{corollary}{Corollary}[section]

\begin{document}

\title{On the multiplicativity of quantum cat maps}

\author{Francesco Mezzadri\footnote[1]{Supported by a Royal
Society Dorothy Hodgkin Fellowship.} \footnote[2]{Permanent
address: Department of Mathematics, University of Bristol,
University Walk,\newline Bristol BS8 1TW, UK.}}

\address{American Institute of Mathematics, 360 Portage Ave, Palo
Alto, CA 94306, USA}

\ead{f.mezzadri@bris.ac.uk}

\mydate{8 April 2002}

\begin{abstract}
The quantum mechanical propagators of the linear automorphisms of
the two-torus (cat maps) determine a projective unitary
representation of the theta group $\Gamma_\theta \subset \SLTZ$.
We prove that there exists an appropriate choice of phases in the
propagators that defines a proper representation of
$\Gamma_\theta$. We also give explicit formulae for the
propagators in this representation.
\end{abstract}

\msc{20C35, 20H05, 81Q50}

\section{Introduction}

The quantization of the linear automorphisms of the two-torus (cat
maps) was first developed by Hannay and Berry~\cite{HB80}, who
were able to determine the quantum propagators for the subgroup of
$\SLTZ$ defined by
\[
 \Gamma_\theta = \left\{A = \Biggl(
\begin{array}{cc}
      a & b \\ c & d
     \end{array} \Biggr) \in \SLTZ \Biggl|\; ab \equiv cd \equiv 0 \bmod 2
     \right \}.
\]
 This set of matrices is often known as the {\it theta
group}. It turns out that it is possible to associate to every
element of $\SLTZ$ a propagator~\cite{Kna90,Esp93}, or quantum
map, which is a unitary operator acting on a finite dimensional
Hilbert space. However, the multiplication among propagators of
different cat maps can be defined only if the quantization is
restricted to certain subgroups of $\SLTZ$, which depend on the
periodicity conditions imposed on the quantum wavefunction (see,
e.g.~\cite{Kna90,Esp93,KMR99}).  It turns out that the theta group
is the largest of these subgroups.

An exact form of Egorov's theorem characterizes the quantum cat
maps, which, as a consequence, are {\it multiplicative} up to a
phase factor, i.e.
\begin{equation}
U_N(AB)=\rme^{2\pi \rmi \zeta(A,B)}U_N(A)U_N(B), \quad A,B \in
\Gamma \subset \SLTZ,
\end{equation}
where $U_N(A)$ denotes the propagator, $\Gamma$ is a subgroup of
$\SLTZ$ and $N$ is the dimension of the Hilbert space in which
$U_N(A)$ acts. In other words, the quantization determines a {\it
projective} representation of $\Gamma$, which is sometimes
referred to as {\it Weil's representation}.

In their original paper, Hannay and Berry conjectured that the
quantization of the cat maps is multiplicative in the theta group,
i.e. it defines a proper representation. The importance of this
property was emphasized by Kurlberg and Rudnick~\cite{KR00}, who
proved that if we restrict to the congruence subgroup defined by
$A \equiv 1_2 \bmod 2$ for $N$ odd and $A \equiv 1_2 \bmod 4$ for
$N$ even, where $1_2$ denotes the $2 \times 2$ identity matrix,
then it is possible to choose the propagators so that the
quantization is multiplicative. The purpose of this paper is to
prove that there exists a choice of phases of the propagators
$U_N(A)$ that determines a proper representation of the whole
theta group. We also give an explicit expression for these phases.

The multiplicativity of the propagators is strictly related to the
existence of a set of unitary operators --- known as Hecke
operators in analogy with a similar phenomenon in the theory of
modular surfaces --- that commute with the map and among
themselves. Indeed, it can be shown that multiplicativity implies
the existence of these symmetries. Most of the mathematical
properties of the quantum cat maps, like the degeneracy in their
spectra, are due to the Hecke operators. Kurlberg and
Rudnick~\cite{KR00} proved that the set of simultaneous
eigenfunctions of a quantum cat map and of its Hecke symmetries
(the Hecke eigenfunctions) become equidistributed in the
semiclassical limit with respect to Liouville measure. They also
obtained rigorous results on the value distribution and extreme
values of a particular class of Hecke eigenfunctions~\cite{KR01}.
Furthermore, the Hecke operators are responsible for spectral
statistics of a significant class of perturbations of the quantum
cat maps that, although being consistent with distributions of
eigenvalues of random matrices, do not belong to the universality
classes expected from the symmetries of the classical
dynamics~\cite{KM00}.

The outline of the article is as follows. In section~\ref{qcm} the
quantum cat maps are introduced and the main results are
presented. The implications of the quantization defined in this
paper on the structure of the Hecke operators are described in
section~\ref{rem}. Section~\ref{alpr} is devoted to the proof of
multiplicativity for the theta group. In sections~\ref{mD2} and
\ref{rmult} we determine a choice of quantum propagators that
defines a proper representation of the theta group. This will also
give an alternative proof of multiplicativity.

\section{The quantum cat maps}
\label{qcm}

The cat maps are the linear automorphisms of the two-torus
$\mathbb{T}^2=\mathbb{R}^2/\mathbb{Z}^2$.  Their dynamics may be
represented by the action of elements of the modular group $\SLTZ$
modulo one. In other words, we consider the symplectic map
\[
\bi{z} \mapsto A \cdot \bi{z}, \quad \bi{z}= \Biggl(
\begin{array}{c} q \\ p \end{array} \Biggr)
 \in \mathbb{T}^2, \quad A \in \SLTZ.
\]
The torus plays the role of phase space, therefore the coordinates
$q$ and $p$ are taken to represent the position variable and its
conjugate momentum.

The action of $A$ on $\mathbb{T}^2$ can be interpreted in terms of
the discrete time evolution of a dynamical system, thus the
corresponding quantum dynamics is determined by a unitary operator
$U_N(A)$. When $\abs{\Tr(A)}> 2$, the dynamics generated by the
classical map is hyperbolic. Since the phase space is compact, the
Hilbert space $\HN$ on which $U_N(A)$ acts is finite dimensional
and may be identified with $\Ltwo(\mathbb{Z}/N\mathbb{Z})$, where
$\mathbb{Z}/N\mathbb{Z}$ is the ring of congruence classes modulo
$N$.  The dimension of $\HN$ and Planck's constant are related via
the condition $2\pi \hbar=1/N$. (We refer the reader to appendix~B
for more details on the quantization of maps on the torus.)

There are many standard ways of mapping classical observables
$f\in C^{\infty}(\mathbb{T}^2)$ into operators $\Op(f)$ acting on
$\HN$. However, elements of $\SLTZ$ cannot be quantized using
these techniques because, although they are symplectic, they
cannot be interpreted as a one-time flow of a Hamiltonian on
$\mathbb{T}^2$~\cite{KMR99}. Therefore, they need an {\it ad hoc}
quantization procedure. Since we are considering linear systems,
it is natural to require that the quantum and classical evolution
commute, or, more precisely, that an exact form of Egorov's
theorem holds:
\begin{equation}
\label{Eg_th}
 U_N(A)^{-1}\Op(f)U_N(A)=\Op(f \circ A), \quad
 f \in C^{\infty}(\mathbb{T}^2).
\end{equation}
An alternative way of understanding~\eref{Eg_th} is the following.
The operator $\Op(f)$ can be expressed as a linear combination of
elements of an irreducible representation $(\tau,\HN)$ with a
given central character $\chi$ of a finite Heisenberg group $H$
(see, e.g.~\cite{KR00}).  Because of the Stone-von Neumann
theorem, there exists a unique isomorphism class of irreducible
representations of $H$ with central character $\chi$. Then, $A \in
\SLTZ$ can be projected into an automorphism of $H$ that fixes the
centre, and thus the map $\tau_A(h)=\tau({}^Ah)$, with $h\in H$,
defines an irreducible representation $(\tau_A, \HN)$ as long as
the two central characters are the same, i.e. $\chi^A=\chi$. These
two representations must be isomorphic. The intertwining operator
$U_N(A):\HN \rightarrow \HN$ from $(\tau,\HN)$ to $(\tau_A,\HN)$
is known as quantum propagator of $A$. Because of Schur's lemma,
$U_N(A)$ is unique up to a phase factor. Many approaches have been
developed to determine
$U_N(A)$~\cite{HB80,Kna90,Esp93,DEG96,Zel97,KR00}, which, because
of~\eref{Eg_th}, are all equivalent up to phase factors.

The topology of the torus constrains the quantum wavefunction to
be periodic up to a phase factor in both the position and momentum
representations, i.e.
\begin{eqnarray*}
\psi(q + m_1) & =  \rme(m_1\varphi_2)\, \psi(q) \\
\hat{\psi}(p + m_2)  & = \rme(-m_2\varphi_1)\, \hat \psi(p), \quad
\boldsymbol{\varphi} =\Biggr( \begin{array}{c}
                                 \varphi_1 \\ \varphi_2
                              \end{array}\Biggl) \in \mathbb{T}^2,
\quad \bi{m} =\Biggr( \begin{array}{c}
                                 m_1 \\ m_2
                      \end{array}\Biggl) \in \mathbb{Z}^2,
\end{eqnarray*}
where $\rme(x) := \rme^{2 \pi \rmi x}$ and
\[
\hat{\psi}(p)=  \frac{1}{\sqrt{2\pi
    \hbar}}\int_{-\infty}^{+\infty} \psi(q) \,
    \rme\left(-\frac{qp}{2 \pi \hbar}\right)\, \rmd q.
\]
Elements of $\SLTZ$ can be quantized only when $\varphi_1$ and
$\varphi_2$ are rational numbers and, given a
$\boldsymbol{\varphi} \in \mathbb{Q}^2/\mathbb{Z}^2$, the
quantization is restricted to a certain subgroup of $\SLTZ$ (see,
e.g.~\cite{Kna90,Esp93,DEG96,KMR99}). When $\boldsymbol{\varphi}=
(0,0)$, then the maps that can be quantized belong to
$\Gamma_\theta$.  These restrictions are due to the constraint
that the central characters $\chi$ and $\chi^A$ of the two
irreducible representations $(\tau,\HN)$ and $(\tau_A,\HN)$ must
be the same. The theta group has index three in $\SLTZ$ and is the
largest subgroup of $\SLTZ$ that can be quantized once the
periodicity conditions have been fixed, in the sense that when
$\boldsymbol{\varphi} \neq (0,0)$ the quantization is restricted
to groups of higher index in $\SLTZ$. In this article we shall
consider only the case when $\boldsymbol{\varphi}=(0,0)$, i.e. the
wavefunction is exactly periodic in both position and momentum
representations.

It is easy to see that Egorov's theorem determines a projective
representation of $\Gamma_\theta$. The aim of this paper is to
prove that it is possible to define a proper representation of
$\Gamma_\theta$. We have the following result.
\begin{theorem}
\label{th1}
 For any positive integer $N$, there exists a choice of phases in
the definition of the propagators $U_N(A)$ such that
\begin{equation}
\label{mult}
 U_N(AB) = U_N(A)U_N(B), \quad A,B \in \Gamma_\theta.
\end{equation}
A choice of propagators that realizes~\eref{mult} is the
following. If
\numparts
\label{operators}
\[
 S^{\pm} = \Biggl( \begin{array}{cc} 0 & \mp 1\\ \pm 1 &
0 \end{array} \Biggr),
\]
then the propagator is the discrete Fourier transform:
\begin{equation}
\label{qft}
 \left[U_N(S^{\pm})\Phi\right](Q) :=
\frac{1}{\sqrt{N}}\sum_{Q' \bmod N} \rme \left(\frac{\pm Q Q'}{N}
\right)\Phi(Q'), \quad \Phi \in \HN.
\end{equation}
The parity matrix
\[
P= \Biggr( \begin{array}{cc} -1 & 0 \\ 0 & -1
\end{array} \Biggl)
\]
has the obvious quantization
\begin{equation}
\label{pmq}
 \left[U_N(P)\Phi\right](Q):= \Phi(-Q), \quad \Phi \in
\HN.
\end{equation}
 If $b = 0$, i.e.
\[
 T_m^{\pm} = \Biggl( \begin{array}{cc} \pm 1 & 0\\ m &
\pm 1 \end{array} \Biggr), \quad m \equiv 0 \bmod 2,
\]
then we have
\begin{equation}
\label{diag_m}
 \left[U_N(T_m^{\pm})\Phi\right](Q) :=
\rme\left(\frac{\pm mQ^2}{2N}\right)\Phi(\pm Q), \quad \Phi \in
\HN.
\end{equation}
Similarly, if $a=0$, i.e.
\[
W_w^{\pm} = \Biggl( \begin{array}{cc} 0 & \pm 1 \\
                    \mp 1 & w \end{array} \Biggl), \quad w \equiv
                    0 \bmod 2,
\]
then we set
\begin{equation}
\label{new_ent}
 \fl \left[U_N(W_w^{\pm })\Phi\right](Q) :=
\frac{1}{\sqrt{N}} \sum_{Q' \bmod N}
 \rme\left(\frac{\pm 1}{2N}\left(wQ^2 - 2 Q Q'
\right)\right)\Phi(Q'), \quad \Phi \in \HN.
\end{equation}
 In all the other cases we define
\begin{eqnarray}
\label{my_cat}
 \fl \left[U_N(A)\Phi\right](Q) :=
\frac{h(a,b)}{\sqrt{N_b}} \sum_{Q' \bmod N}
G\left(\alpha,\beta,\gamma(Q,Q')\right) \nonumber \\
\quad \times \rme\left(\frac{1}{2Nb}\left(dQ^2 - 2 Q Q'  + a
{Q'}^2\right)\right)\Phi(Q'), \quad \Phi \in \HN,
\end{eqnarray}
\endnumparts
where $h(a,b)$ and $G(\alpha,\beta,\gamma)$ are defined in
equations~\eref{hf} and~\eref{GS} respectively.
\nonumber
\end{theorem}
Formula~\eref{my_cat} needs a few words of explanation.  Firstly,
we have
\begin{equation}
\label{hf}
h(a,b) :=
\cases{\Jsym{\abs{a}}{\abs{b}}\rme\left(\sign(ab)\left(\abs{b} -
1\right)/8\right) & if $a$ is even, \\
\Jsym{\abs{b}}{\abs{a}}\rme\left(-\sign(ab)\abs{a}/8\right) & if
$a$ is odd,}
\end{equation}
where $\Jsym{p}{q}$ is the Jacobi symbol (see appendix A for the
definition and properties of the Jacobi symbol) and $\sign(x)$ is
the sign function
\[
\sign(x):=\cases{1 & {\rm if $x> 0$,} \\
       -1 & {\rm if $x< 0$,} \\
       0 & {\rm if $x=0$.}}
\]
 The function $h(a,b)$ is a phase factor which distinguishes
definition~\eref{my_cat} from the quantization that Hannay and
Berry introduced in their original article~\cite{HB80}.  More
precisely, Hannay and Berry's propagator is $U_N^{{\rm HB}}(A)=
\sqrt{i}U_N(A)/h(a,b)$.

The term $G\left(\alpha,\beta,\gamma(Q,Q')\right)$ is a normalized
Gauss sum:
\begin{equation}
\label{GS} G(\alpha,\beta,\gamma(Q,Q')) :=
\frac{1}{\sqrt{\abs{\beta}}}\sum_{k \bmod
\abs{\beta}}\rme\left(\frac{1}{2\beta}\left(\alpha k^2 +
\gamma(Q,Q') k\right)\right),
\end{equation}
where $\alpha := N_b a$, $\beta := b'$, $\gamma := 2\left(aQ' -
Q\right)/(b,N)$ and $(b,N)$ is the greatest common divisor of $b$
and $N$, which we shall always take to be positive.  In the
previous definitions, in~\eref{my_cat} and in what follows we use
the notation
\begin{eqnarray*}
N_a := N/(a,N), & a' := a/(a,N),\\
N_b := N/(b,N), & b':=b/(b,N),\\
N_d := N/(d,N), & d':=d/(d,N), \\
N_{ab} := N/((a,N)(b,N)),\qquad & N_{bd}:=N/((b,N)(d,N)).
\end{eqnarray*}
The sum~\eref{GS} is different from zero only if $\alpha$ and
$\beta$ are coprime integers, $\gamma$ is also an integer and
$\alpha \beta + \gamma$ is even. $G(\alpha,\beta,\gamma)$ can be
explicitly computed:
\begin{equation}
\label{evgaussum}
 \fl G(\alpha,\beta,\gamma) =  \left \{\begin{array}{l}
\Jsym{\abs{\alpha}}{\abs{\beta}} \rme
  \left(-\sign(\alpha\beta)\left(\abs{\beta}-1\right)/8\right)
  \rme\left(-\frac{\alpha \invint{\alpha}{\abs{\beta}}^2}{2\beta}
  \left(\frac{\gamma}{2}\right)^2\right)  \\
\hbox{\rm if $\alpha$ is even, $\beta$ odd and $\gamma$ even,} \\
\Jsym{\abs{\beta}}{\abs{\alpha}} \rme
  \left(\sign(\alpha\beta)\abs{\alpha}/8\right)
  \rme\left(-\frac{\alpha \invint{\alpha}{\abs{\beta}}^2}{2\beta}
  \left(\frac{\gamma}{2}\right)^2\right)  \\
\hbox{\rm if $\alpha$ is odd, $\beta$ even and $\gamma$ even,}\\
\Jsym{\abs{\alpha}}{\abs{\beta}} \rme
  \left(-\sign(\alpha\beta)\left(\abs{\beta}-1\right)/8\right)
  \rme\left(-2\alpha \invint{4\alpha}{\abs{\beta}}^2\gamma^2/\beta\right)\\
  \hbox{\rm if $\alpha$ is odd, $\beta$ odd and $\gamma$ odd.}
 \end{array} \right.
\end{equation}
Here $\invint{p}{q}$ denotes the inverse integer of $p$ modulo
$q$, where $p$ and $q$ are mutually prime, i.e. the only integer
modulo $q$ such that
\[
p\invint{p}{q} \equiv 1 \bmod q.
\]
Moreover, the Euler-Fermat theorem gives
\[
\invint{p}{q} \equiv p^{\phi(q) - 1} \bmod q,
\]
where $\phi(q)$ is Euler's function, which is defined as the
number of integers less than and mutually prime to $q$. (For a
detailed explanation of~\eref{GS} and~\eref{evgaussum} we refer
the reader to~\cite{HB80}.)

It was pointed out by Hannay and Berry~\cite{HB80} that the Gauss
sum~\eref{GS} is invariant if
\begin{equation}
\label{sub}
 N_ba \rightarrow N_bd, \quad
\frac{2}{(b,N)}\left(aQ' - Q\right) \rightarrow
\frac{2}{(b,N)}\left(dQ - Q'\right).
\end{equation}
The reason is quite simple: if $f(Q)$ is a function defined on
$\mathbb{Z}/m\mathbb{Z}$, then the sum $\sum_{Q \bmod m} f(Q)$ is
invariant if $Q \rightarrow lQ$, where $(l,m)=1$.  The
substitutions~\eref{sub} are equivalent to replacing $k$ by $-dk$
in~\eref{GS}. Similarly, we have the equality
\begin{equation}
\label{invp_h}
 h(a,b)=h(d,b),
\end{equation}
which is proven in appendix C.

It is straightforward to check that the
operators~\eref{qft},~\eref{pmq},~\eref{diag_m} and~\eref{new_ent}
satisfy~\eref{Eg_th}; it was shown by Knabe~\cite{Kna90} and Degli
Esposti~\cite{Esp93} that the propagator introduced by Hannay and
Berry~\cite{HB80}, and therefore also definition~\eref{my_cat},
obeys an exact form of Egorov's theorem.  Thus, the quantization
procedure that we gave defines a projective representation of
$\Gamma_\theta$. In the following sections we will prove that it
is proper representation. In section~\ref{rem} we will show that
the definitions of the propagators (2.3) depend only on the
reduction of $A$ modulo $4N$. Therefore, since the projection
\[
\pi: \SLTZ \rightarrow {\rm SL}(2,\mathbb{Z}/4N\mathbb{Z})
\]
is surjective (see, e.g.~\cite{Shi71}), formulae (2.3) also define
a representation of the group
\[
\Gamma_\theta(4N) = \left\{A = \Biggl(
\begin{array}{cc}
      a & b \\ c & d
     \end{array} \Biggr) \in {\rm SL}(2,\mathbb{Z}/4N\mathbb{Z})
      \Biggl|\; ab \equiv cd \equiv 0 \bmod 2
     \right \}.
\]

There exist partial results in the direction of the first part of
theorem~\ref{th1}.  The first goes back to Schur~\cite{Shu07}, who
proved that when $p$ is an odd prime, any projective
representation of ${\rm SL}(2,\mathbb{Z}/p \,\mathbb{Z})$ can be
modified to give a representation.  An analogous result was
obtained in~\cite{Nob76,BI86}. More generally, the same property
holds for ${\rm SL}(2,\mathbb{Z}/m\, \mathbb{Z})$ when $m
\notequiv 0 \bmod 4$~\cite{Men67,Bey86}. Kurlberg and
Rudnick~\cite{KR00} proved that there exists a choice of
propagators that defines a representation of the congruence
subgroup
\[
 \Gamma(4,2N) = \left\{A
 \in {\rm SL}(2,\mathbb{Z}/2N\mathbb{Z})
 \left |\; \cases{A \equiv 1_2  \bmod 2& for $N$ odd \\
                            A \equiv 1_2 \bmod 4 & for $N$ even}
                             \right. \right \}.
\]
Similar results can be found in~\cite{Kub69,Gel76}.

\section{The Hecke operators}
\label{rem} The quantum cat maps are characterized by the
existence of a group of commutative unitary symmetries known as
Hecke operators. As mentioned in the introduction, these
symmetries are responsible of most of the arithmetical properties
of the quantum cat maps. Since theorem~\ref{th1} has important
consequence on the structure of such operators, we briefly
introduce them and describe how they are related to the
propagators defined in formulae~(2.3). The proof of
theorem~\ref{th1} will be presented in the following sections.

If two cat maps are equivalent modulo $2N$, then their quantum
propagators differ by a phase factor.  This property was already
discovered by Hannay and Berry~\cite{HB80} and is a direct
consequence of formulae~\eref{evgaussum}.  Kurlberg and
Rudnick's~\cite{KR00} quantization is a map
\begin{equation}
\label{rmap}
 \rho :{\rm SL}(2,\mathbb{Z}/2N\mathbb{Z}) \rightarrow
{\rm U}(N), \quad A \mapsto U_N^{{\rm KR}}(A).
\end{equation}
So by construction, if $A \equiv B \bmod 2N$ then $U_N^{{\rm KR}}
(A)=U_N^{{\rm KR}}(B)$. Thus, multiplicativity implies the
existence of Hecke operators. In fact, suppose that $AB \equiv BA
\bmod 2N$, then we have
\[
U_N^{{\rm KR}}(AB)=U_N^{{\rm KR}}(A)U_N^{{\rm KR}}(B) =U_N^{{\rm
KR}} (B)U_N^{{\rm KR}}(A)=U_N^{{\rm KR}}(BA),
\]
i.e. $U_N^{{\rm KR}}(B)$ is a symmetry of $U_N^{{\rm KR}}(A)$,
even though $A$ and $B$ do not commute exactly.

Kurlberg and Rudnick~\cite{KR00} proved that the map~\eref{rmap}
can be defined in such a way that $\rho$ is a proper
representation when restricted to the congruence subgroup
$\Gamma(4,2N)$. They also proved that given $A \in \Gamma(4,2N)$
the number of elements of $\Gamma(4,2N)$ that commute with $A$ and
among themselves is of order $N$.

We conclude by showing that the propagators defined in formulae
(2.3) depend only on the reduction of $A$ modulo $4N$ and
therefore they define a proper representation of
$\Gamma_\theta(4N)$. As a consequence, in this case the Hecke
symmetries of a quantum cat map $U_N(A)$ are those $U_N(B)$ such
that
\[
AB \equiv BA \bmod 4N.
\]
The order of the group of Hecke symmetries of a given $U_N(A)$ is
of order $N$ also in this case. This follows directly from
Kurlberg and Rudnick's result, which implies that the number of
equivalence classes of matrices modulo $4N$ in
\[
\Gamma(4)=\{A \in \SLTZ |\, A \equiv 1_2 \bmod 4 \}
\]
that commute modulo $4N$ among themselves and with a given $A$ is
of order $N$. Since $ \Gamma(4)$ is of finite order in
$\Gamma_\theta$, the same statement holds in $\Gamma_\theta$. It
is worth noting that the reduction modulo $4N$ in this respect is
essential, because the only matrices that commute exactly with $A
\in \SLTZ$ are $P$ and the powers of the primitive matrix of
$A$~\footnote{$A_0 \in \SLTZ$ is primitive if $A_0=A^k$, $A\in
\SLTZ$, implies $k=1$.}.

It also is important to point out that if there exists a choice of
phases in the definition of the propagators such that
\[
U_N(A)U_N(B) = U_N(B)U_N(A)
\]
then $U_N(A)$ and $U_N(B)$ commute for any choice; this is not the
case with the multiplicativity property~\eref{mult}.

Finally, we have the following.
\begin{corollary}
\label{cor1}
 The propagators $U_N(A)$ defined in formulae (2.3)
depend only on the reduction of $A$ modulo $4N$.
\end{corollary}
\begin{proof}
It is trivial to see that the propagators~\eref{diag_m}
and~\eref{new_ent} are invariant if $m \mapsto m + 4Nk$ and $w
\mapsto w + 4Nk'$. Now, suppose that $A \equiv B \bmod 4N$ and
$U_N(A)$ and $U_N(B)$ are of the form~\eref{my_cat}. We have
\begin{equation}
\label{mod_red}
 A = B + 4NM =B(1_2 + 4NB^{-1}M),
\end{equation}
where $M$ is a matrix with integer entries.  The propagators of
both sides of~\eref{mod_red} are
\[
U_N(A)=U_N(BC)=U_N(B)U_N(C),
\]
where $C=1_2 +4NB^{-1}M$.  We know that
\[
[U_N(C)\Phi](Q)=\rme(\mu)\Phi(Q), \quad \Phi \in \HN.
\]
The phase factor $\rme(\mu)$ can be determined directly
from~\eref{my_cat} and~\eref{hf}:
\[
\fl
\rme(\mu)=h(a,b)\Jsym{\abs{b}/N}{\abs{a}}\rme(\sign(ab)\abs{a}/8)=
\Jsym{N}{\abs{a}} , \quad C=\Biggl(\begin{array}{cc} a & b
\\c & d
\end{array} \Biggr).
\]
Now, since $\abs{a}= \pm 1 \bmod 4N$, from
property~\eref{mod-eight} of the Jacobi symbol we have
\[
\Jsym{N}{\abs{a}}=1.
\]
\end{proof}
More generally, using~\eref{evgaussum} one can show that if
\[
A \equiv \Biggl(\begin{array}{cc} \invint{a}{4N} & 0 \\ 0 & a
\end{array}\Biggl) \bmod 4N, \quad (a,4N)=1,
\]
then
\[
[U_N(A)\Phi](Q)=\Jsym{N}{\abs{a}}\Phi(aQ), \quad \Phi \in \HN.
\]
Corollary~\ref{cor1} then follows immediately. It is also
important to point out that if $A \equiv B \bmod 2N$, then
\[
U_N(A)=\Jsym{N}{\abs{a}}U_N(B).
\]
However, since $\abs{a}=\pm 1 \bmod 2N$, the Jacobi symbol
$\Jsym{N}{\abs{a}}$ is not necessarily one.

\section{Relations and multiplicativity for $\Gamma_\theta$}
\label{alpr}
 Given an abstract group $\Gamma$, it is always possible to
choose a subset
\[
G = \{g_1,g_2,\ldots,g_n,\dots\}, \quad G \subset \Gamma
\]
such that each element of $\Gamma$ can be written as product, or
string, of a finite number of $g_i$, i.e. $\forall h \in \Gamma$,
$h=a_1a_2 \cdots a_r$, where $a_j=g_i^\epsilon$ and $\epsilon =
\pm 1$. We call $G$ a set of generators.  The cardinality of $G$
may be finite or infinite; the groups that we are concerned with
are all finitely generated.

It is clear that if we choose appropriately a set of generators of
$\Gamma_\theta$ and then we understand how their propagators
behave, we may be able to make some progress on the
multiplicativity properties of arbitrary elements of
$\Gamma_\theta$.  Indeed, knowing a set of generators, say
$G=\{g_1,g_2,\ldots,g_n \}$, one might be tempted to think that if
the $U_N(g_i)$ obey~\eref{Eg_th} and we define
\begin{equation}
\label{mdef}
U_N(A)=U_N(a_1\cdots a_k)=U_N(a_1) \cdots U_N(a_k),
\quad a_j =g_i^\epsilon, \quad \epsilon = \pm 1,
\end{equation}
then multiplicativity would be automatic. This would be the case
if $\Gamma_\theta$ were a {\it free} group, i.e. if no relations
existed among the generators or, in other words, if each element
of $\Gamma_\theta$ could be written in a unique way as a finite
product $a_1\cdots a_k$\footnote[1]{More precisely, this statement
is true only for {\it reduced words}, i.e. words where no pair
$a_ja_{j+1}$, $j=1,\ldots, k-1$, is of the form $g_i^\epsilon
g_i^{-\epsilon}$. }, each $a_j$ being some $g_i^\epsilon$, where
$\epsilon = \pm 1$. However, in general a given $A \in
\Gamma_\theta$ can be written in many different ways as product of
generators and one must check that such relations hold among the
$U_N(g_i)$ too.

The main result of this section is the following.
\begin{theorem}
\label{th2}
The group $\Gamma_\theta$ is generated by the set of
matrices $G=\{S^+,P,T^+_2\}$, where
\begin{equation}
\label{generators}
S^+ = \Biggr(\begin{array}{cc} 0 & -1 \\ 1 & 0
\end{array}\Biggl), \quad
P= \Biggl(\begin{array}{cc} -1 & 0 \\ 0 & -1 \end{array}\Biggr),
\quad
 T_2^+= \Biggl(\begin{array}{cc} 1 & 0 \\ 2 & 1 \end{array}
 \Biggr).
\end{equation}
The only relations among $S^+,P,T^+_2$ are
\begin{equation}
\label{rel} P^2=1_2, \quad  {S^+}^2P=1_2, \quad S^+P=PS^+, \quad
T^+_2P=PT^+_2.
\end{equation}
Furthermore, if $U_N(S^+)$, $U_N(P)$ and $U_N(T^+_2)$ are given
by~\eref{qft}, \eref{pmq} and \eref{diag_m}, then the following
relations hold:
\begin{equation}
\label{relq}
\begin{array}{l}
U_N(P)^2 = {\rm 1}_N, \\
U_N(S^+)^2U_N(P)= {\rm 1}_N, \\
U_N(S^+)U_N(P)=U_N(P)U_N(S^+),\\
 U_N(T^+_2)U_N(P)=U_N(P)U_N(T^+_2).
 \end{array}
\end{equation}
\end{theorem}
This theorem automatically gives $U_N(AB)=U_N(A)U_N(B)$ once
$U_N(A)$ is defined according to~\eref{mdef}.   However, $U_N(A)$
is given implicitly as product of propagators of generators and
this may be inconvenient in applications. We shall address this
issue in sections~\ref{mD2} and \ref{rmult}.

The statement that $G=\{S^+,P,T^+_2 \}$ is a set of generators of
$\Gamma_\theta$ with relations~\eref{rel} is a direct consequence
of well known results in the theory of modular forms (see,
e.g.~\cite{Ran76}). However, since it is fundamental to our work,
we shall provide a complete proof. The last part of the theorem
can be easily checked by direct multiplication.

We begin by proving that $S^+$ and $T^+_2$ generate
$\Gamma_\theta$.  Then, since ${S^+}^2 = P$, $G$ is also a set of
generators.
\begin{lemma}
The matrices $T_2^+$ and $S^+$ are a set of generators of
$\Gamma_\theta$.
\end{lemma}

\begin{proof}
The proof consists of showing that by successive multiplications
from the right by $T^+_{\pm 2}$ and $S^{\pm}$ an arbitrary matrix
$A \in \Gamma_\theta$ can be reduced to either $T_{\pm 2}^+$ or
$S^{\pm}$.

Let us consider first the two cases when $b=0$ and $b=\pm 1$.
Firstly, suppose that $b=0$, i.e. $A = T_m^{\pm}$ but $m \neq \pm
2$. If $A=T_m^-$ then by applying $P={S^+}^2$ we obtain
$T_{-m}^+$. Clearly, $A$ can be reduced to $T_{\pm 2}^+$ by
repeated multiplications from the right by $T_{\pm 2}$. If,
instead, $b = \pm 1$, repeated applications of $T_{\pm 2}^+$ will
reduce $A$ either to $S^{\pm}$ or to a matrix of the form
\begin{equation}
\label{n_form} W^\pm_w = \Biggl( \begin{array}{cc} 0 & \pm 1 \\
\mp 1 & w
\end{array} \Biggr), \quad w \equiv 0 \bmod 2.
\end{equation}
By multiplying~\eref{n_form} from the right by $S^{\pm}$, the
resulting matrix will have $b = 0$.

Suppose now that $\abs{b} > 1$. If $\abs{a} < \abs{b}$, we
multiply $A$ from the right by $S^{\pm}$, so that $\abs{a} >
\abs{b}$; next we apply appropriately $T_{\pm 2}^+$ until $\abs{a}
< \abs{b}$; we then apply $S^{\pm}$ and repeatedly multiply $A$ by
$T_{\pm 2}^+$ until $\abs{a} < \abs{b}$, and so on. Since the
elements of $A$ are integers, this procedure will stop after a
finite number of steps when $a=0$. Then $A$ is of the
form~\eref{n_form}.
\end{proof}

The relations~\eref{rel} can be verified straightforwardly by
matrix multiplication. To complete the proof of theorem~\ref{th2}
we only have to check that no other relations independent
of~\eref{rel} exist, i.e. if $W = a_1a_2 \cdots a_k =1_2$, where
either $a_j \in G$ or $a_j^{-1}\in G$, then $W$ can be mapped into
the void word by repeated applications of~\eref{rel}.

Let us consider the quotient group $\PT= \Gamma_\theta/\Lambda$,
where $\Lambda=\{1_2,P\}$ is the centre of $\Gamma_\theta$ and the
kernel of the homomorphism $\sigma: \Gamma_\theta \rightarrow
\PT$. Let us denote
\begin{equation}
\label{genPT}
 S = \sigma(S^\pm), \qquad T_2=\sigma(T^\pm_2).
\end{equation}
Clearly $S$ and $T_2$ generate $\PT$ and $\sigma$ maps the
relations~\eref{rel} into  $S^2 =1$. We now proceed in two steps:
firstly, we prove that
\begin{equation}
\label{relptb}
 \PT = \left \langle S, T_2 \, \biggl | S^2 =
 1 \right \rangle.
\end{equation}
Secondly, using~\eref{relptb}, we show that~\eref{rel} are the
only relations in $\Gamma_\theta$.

Let us introduce the groups
\[
\Gamma(2)= \{A \in \SLTZ |\, A \equiv 1_2 \bmod 2 \} \quad {\rm
and } \quad \PGT = \Gamma(2)/\Lambda.
\]
It turns out  (see, e.g.~\cite{Ran76}) that $\PGT$ is a free
normal subgroup of $\PT$ of index two generated by
\begin{equation}
\label{gpg2}
 T_2 \quad {\rm and} \quad \overline{T}_2=ST^{-1}_2S.
\end{equation}
Furthermore, we have
\begin{equation}
\label{div}
 \PT = \PGT + S\, \PGT.
\end{equation}
Because of~\eref{gpg2}, proving~\eref{relptb} is equivalent to
showing that
\begin{equation}
\label{srel}
 \PT = \left \langle S, T_2, \overline{T}_2 \biggl | S^2 = {\rm
1}, ST_2S\overline{T}_2={\rm 1} \right \rangle.
\end{equation}
Now, we have $\PT=\langle S, T_2, \overline{T}_2\rangle$.
Therefore, by~\eref{div} each word
\begin{equation}
\label{rw}
 W = a_1a_2\cdots a_k
\end{equation}
of the symbols $S^\epsilon$, $T_2^\epsilon$ and
$\overline{T}_2^{\, \epsilon}$, $\epsilon = \pm 1$, can be mapped
by a certain set of relations in $\PT$ into a string
\begin{equation}
W'=a_1'a_2'\cdots a_r',
\end{equation}
where either $W'$ is a sequence of the symbols $T_2^\epsilon$ and
$\overline{T}_2^{\, \epsilon}$ or a word where $a_1'=S^\epsilon$
and $a_2'a_3'\cdots a_r'$ is a string that does not contain
$S^{\epsilon}$. Since $\PGT$ is a free group, there are no
relations between $T_2$ and $\overline{T}_2$,
therefore~\eref{srel} and~\eref{relptb} are a consequence of the
following.
\begin{lemma}
 The map $W \mapsto W'$ is achieved by the relations
\begin{equation}
\label{srela}
 S^2 = 1, \quad ST_2S\overline{T}_2 = {\rm 1}.
\end{equation}
\end{lemma}
\begin{proof}
Using $S^2={\rm 1}$ any word $W$ can be transformed into a string
where $S$ appears only with exponent one.  Next, using both
relations~\eref{srela}, all the sequences of the form
\begin{equation}
\label{fel} \cdots ST_2^{\epsilon_m \,m}S\cdots, \quad \cdots
S\overline{T}^{\, \epsilon_n \, n}_2S  \cdots, \quad \epsilon_m,
\epsilon_n = \pm 1, \quad m,n \in \mathbb{Z}^+
\end{equation}
can be turned into powers of $T_2$ and $\overline{T}_2$.  Suppose
now that only an even number of $S$s are left into $W$. Then, we
break $W$ into strings of the form
\begin{equation}
\label{sel} \cdots ST_2^{\epsilon_m \,m}\overline{T}_2^{\,
\epsilon_n \, n} \cdots T_2^{\epsilon_p p}S\cdots, \quad
\epsilon_m,\epsilon_m, \epsilon_p = \pm 1, \quad m,n,p \in
\mathbb{Z}^+.
\end{equation}
Such sequences can be transformed into powers of $T_2$ and
$\overline{T}_2$ by inserting $S^2$ between each pair
$T_2^{\epsilon_m \,m}\overline{T}_2^{\, \epsilon_n \, n}$.  The
resulting word $W'$ does not contain $S$. Instead, if $W$ contains
an odd number of $S$s, the transformation~\eref{sel} will leave
only one $S$ in $W$. Now, either $S$ is the first symbol, in which
case we stop, or $W$ begins with a string of the type
\[
T_2^{\epsilon_m \,m}\overline{T}_2^{\, \epsilon_n \, n} \cdots
T_2^{\epsilon_p \, p}S\cdots, \quad
\epsilon_m,\epsilon_n,\epsilon_p = \pm 1, \quad m,n,p \in
\mathbb{Z}^+.
\]
The symbol $S$ can then be moved in the first position by
multiplying the word by $S^2$ from the left and by inserting $S^2$
between each pair $T_2^{\epsilon_m \,m}\overline{T}^{\, \epsilon_n
\, n}_2$.
\end{proof}
We now have left to show that no other relations except~\eref{rel}
exist among $S^+$, $P$ and $T^+_2$. Since $P$ commutes with both
$S^+$ and $T^+_2$, the only other possible relations must be of
the form
\numparts
\begin{eqnarray}
\label{prel1}
{T_2^+}^{\epsilon_m m}P= 1_2, \quad \epsilon_m = \pm 1, \quad m \in \mathbb{Z}^+\\
\label{prel2}
Pa_1a_2\cdots a_k =1_2,\\
\label{prel3}
 a_1'a_2'\cdots a_r' =1_2,
\end{eqnarray}
\endnumparts
where $a_1a_2\cdots a_k$ and $a_1'a_2'\cdots a_k'$ are strings
that contain both $S^+$ and $T^+_2$ but not $P$. Now, $T^+_2$ is
of infinite order; this excludes relation~\eref{prel1} because,
together with $P^2=1_2$, it would imply that $T_2^+$ is of finite
order.  Relations of the type~\eref{prel2} and~\eref{prel3} are
excluded because their images under the homomorphism
$\sigma:\Gamma_\theta \rightarrow \PT$ would be relations in $\PT$
involving $S$ and $T_2$.  This completes the proof of
theorem~\ref{th2}.

\section{The choice of propagators}
\label{mD2}

Theorem~\ref{th2} says that there exists a choice of phases in the
definition of the propagators such that
\begin{equation}
\label{smult}
 U_N(AB)=U_N(A)U_N(B), \quad A,B \in \Gamma_\theta.
\end{equation}
The quantum map $U_N(A)$ for arbitrary $A \in \Gamma_\theta$ is
then given as a product of a certain finite sequence of
$U_N(T^+_{\pm 2})$ and $U_N(S^{\pm})$. Our goal is to prove that
the propagators~(2.3) are equivalent to such a product. This will
be the aim of this section and section~\ref{rmult}.

Our strategy will be to prove that multiplicativity holds if any
of the propagators (2.3) is multiplied either from the right or
from the left by any of $U_N(T_{\pm 2}^+)$ or $U_N(S^{\pm})$. This
will imply the equivalence of definitions (2.3) and~\eref{mdef},
and will provide an independent proof of~\eref{smult}.

It is a straightforward exercise to check that if we multiply any
of the operators \eref{qft}, \eref{pmq}, \eref{diag_m} and
\eref{new_ent} from the left and from the right by $U_N(T_{\pm
2}^+)$ and $U_N(S^{\pm})$, then multiplicativity holds. Thus, it
remains to be proved for the subset of $\Gamma_\theta$ such that
$U_N(A)$ is of the form~\eref{my_cat}, i.e.
\[
M = \left \{A \in \Gamma_\theta \mid a \neq 0, b \neq 0\right \}.
\]
The product of $A \in M$ with one of the generators of
$\Gamma_\theta$ may not belong to $M$. However, it turns out that
we need to check multiplicativity only if the product of $A$ with
a given generator still belongs to $M$.  For example, let us
suppose that $AT^{\pm}_m = S^{\pm}$. Then we have
\[
U_N(AT^{\pm}_{m})U_N(T^{\pm}_{-m}) = U_N(S^{\pm})U_N(T^{\pm}_{-m})
= U_N(A),
\]
which implies
\[
U_N(AT^{\pm}_{m}) = U_N(A)U_N(T^{\pm}_{m}).
\]

Moreover, it turns out that if multiplicativity holds if $A$ is
multiplied from the right by the generators of $\Gamma_\theta$,
then it also true when $A$ is multiplied from the left and vice
versa. For example, let us consider $S^\pm$. Since
$(S^{\pm}A)^{-1} = A^{-1}S^{\mp}$ and  $U_N(A^{-1})=U_N(A)^{-1}$,
which can be verified directly from definitions (2.3), we obtain
\[
U_N(S^{\pm}A)^{-1}=U_N( A^{-1}S^{\mp})=
 U_N(A^{-1})U_N(S^{\mp})=(U_N(S^{\pm})U_N(A))^{-1},
\]
which implies
\[
U_N(S^{\pm}A)=U_N(S^{\pm})U_N(A).
\]
An analogous argument applies to $T^+_{\pm 2}$.

Proving multiplicativity for $T^+_{\pm 2}$ is quite
straightforward. We need to show that
\begin{equation}
\label{mt2}
 U_N(T_{\pm 2}^+A) = U_N(T_{\pm 2}^+)U_N(A), \quad A,T_{\pm 2}^+A \in M.
\end{equation}
We now apply both sides of the above equation to the Kronecker
delta function $\delta_0(Q)$ (see appendix B for the definition of
$\delta_\nu(Q)$) and evaluate their images at $Q = 0$.  Since the
first row of $A$ is not affected by the multiplication by
$T^+_{\pm 2}$, we immediately obtain
\numparts
\begin{equation}
\label{mt2a}
 \left[U_N(T_{\pm 2}^+A)
 \delta_0\right](0)=\sqrt{(b,N)}h(a,b)G(N_ba,b',0)
\end{equation}
and
\begin{equation}
\label{mt2b}
 \left[U_N(T_{\pm 2}^+)U_N(A)
 \delta_0\right](0)=\sqrt{(b,N)}h(a,b)G(N_ba,b',0).
\end{equation}
\endnumparts
Since the propagators $U_N(A)$ form a projective representation of
$\Gamma_\theta$, equations (5.3) imply~\eref{mt2}.

Proving multiplicativity when the generators are $S^{\pm}$ is more
involved and will be dealt with in the next section.

\section{Multiplication by $S^{\pm}$}
\label{rmult}
We now show that
\begin{equation}
\label{mult_t2}
 U_N(AS^{\pm}) = U_N(A)U_N(S^{\pm}), \quad A,AS^{\pm} \in M.
\end{equation}
We shall discuss the proof of~\eref{mult_t2} only for $S^+$, since
\[
AS^-=-AS^+, \quad A,AS^- \in M.
\]
Let us consider the two disjoint subsets
\begin{eqnarray*}
K_{\rm R} & = \left\{ A \in M \mid A \equiv S^+ \bmod 2, \quad
 AS^+\in M \right\}, \\
L_{\rm R} & = \left\{ A \in M \mid A \equiv 1_2 \bmod 2, \quad
AS^+ \in M \right \}.
\end{eqnarray*}
Clearly $K_{\rm R}=L_{\rm R}S^+$ and $L_{\rm R}= K_{\rm R}S^+$.
Thus, if~\eref{mult_t2} holds when $A \in K_{\rm R}$, it is also
true if $A \in L_{\rm R}$. In fact, suppose that $A \in L_{\rm
R}$, then we have $B=AS^+ \in K_{\rm R}$ and
\[
U_N(A)U_N(S^+) =U_N(BS^-)U_N(S^+)=U_N(B)=U_N(AS^+).
\]
Therefore, we need to prove~\eref{mult_t2} only for $A \in K_{\rm
R}$.

As in the previous section, we shall check multiplicativity by
applying both sides of equation~\eref{mult_t2} to $\delta_\nu(Q)$.
The integer $\nu$ and the value of $Q$ at which the image of
$\delta_\nu(Q)$ will be evaluated will be chosen appropriately in
order to make the algebra simple.

We have three different cases to consider:
\begin{itemize}
\item $N_a$ is odd and $a'$ is even;
\item $N_a$ is even and $a'$ is odd;
\item $N_a$ and $a'$ are both odd.
\end{itemize}

\subsection{Case 1: $N_a$ odd and $a'$ even}
\label{c1r}
 The appropriate function to use in this case is
$\delta_0(Q)$. Let us first apply it to the left-hand side
of~\eref{mult_t2}. We easily obtain
\begin{equation}
\label{simsum} \fl \left[U_N(AS^+)\delta_0\right](0) =
\sqrt{(a,N)}
\Jsym{\abs{a}}{\abs{b}}\Jsym{\abs{a'}}{N_{ab}\abs{b'}}
\rme\left(\sign(ab)\left(\abs{b} - N_{ab}\abs{b'}
\right)/8\right),
\end{equation}
where we have used $(b,N)^2 \equiv 1 \bmod 8$.

The term $\left[U_N(A)U_N(S^+)\delta_0\right](0)$ needs some more
work. The action of $U(A)$ on a vector $\Phi \in \HN$ is given by
\begin{eqnarray}
\label{comp_ex} \fl \left[ U_N(A) \Phi \right](Q) = \frac{1}{\sqrt
N_b} \Jsym{\abs{a}}{\abs{b}}\Jsym{N_{ab} \abs{a'}}{\abs{b}'}
\rme\left(\sign (ab)
\left(\abs{b} - \abs{b'}\right)/8\right) \nonumber \\
\times \sum_{\substack{Q' \bmod N \cr (b,N) \mid (aQ' - \; Q)}}
\rme\left(\frac{1}{2N b} \left(
d Q^2 - 2QQ' + a{Q'}^2 \right)\right) \nonumber \\
\times \rme \left( -  \frac{a N_b \invint{\abs{a}
N_b}{\abs{b'}}^2}{2b'} \left(\frac{aQ' - Q}{(b,N)} \right)^2
\right)\Phi(Q').
\end{eqnarray}

The condition $ad - bc = 1$ implies that $a$ and $b$ are mutually
prime and that $d \equiv \invint{a}{\abs{b}} \bmod \abs{b}$.
Moreover, we have
\numparts
\begin{eqnarray}
 \label{inv_intg1}
 \invint{a N_b}{\abs{b'}} \equiv  d
 N_b^{(\phi(\abs{b'}) -1)} \bmod \abs{b'},\\
 \label{inv_intg2}
 \invint{b'}{N_b} \equiv \frac{1 - N_b^{2\phi(\abs{b'})}}{b'}
 \equiv \frac{1 - N_b^{\phi(\abs{b'})}}{b'} \bmod N_b.
\end{eqnarray}
\endnumparts
Using the above congruences and replacing $\Phi(Q)$ by
$\left[U_N(S^+)\delta_0\right](Q)$, equation~\eref{comp_ex}
becomes
\begin{eqnarray}
\label{finsum}
 \fl \left[U_N(A)U_N(S^+)\delta_0\right](0)  =
 \frac{1}{\sqrt N_b}
\Jsym{\abs{a}}{\abs{b}}\Jsym{N_{ab} \abs{a'}}{\abs{b'}}
\rme\left(\sign (ab)
\left(\abs{b} - \abs{b'}\right)/8\right) \nonumber \\
 \times \sum_{Q \bmod N_b } \rme\left(\frac{1}{2N_{ab}}
 \invint{b'}{N_b} a'Q^2 \right).
\end{eqnarray}
Since the above Gauss sum does not depend on the choice of the
representative in the equivalence class of $\invint{b'}{N_b}$, we
can choose $\sign \bigl[\invint{b'}{N_b} \bigr]=- \sign(b)$. This
is also consistent with~\eref{inv_intg2}. The sum in~\eref{finsum}
now becomes
\begin{eqnarray}
\label{finsume}
 \fl \left[U_N(A)U_N(S^+)\delta_0\right](0) =
 \sqrt{(a,N)}\Jsym{\abs{a}}{\abs{b}}\Jsym{N_{ab}\abs{a'}}{\abs{b'}}
 \Jsym{-\sign(b)\abs{a'}\invint{b'}{N_b}}{N_{ab}}\nonumber \\
\quad \times \rme\left(\sign (ab)%
 \left(\abs{b}-\abs{b'} + N_{ab} - 1\right)/8\right).
\end{eqnarray}
The above equation is then transformed into~\eref{simsum} by using
the equality
\begin{eqnarray*}
\fl\Jsym{N_{ab}\abs{a'}}{\abs{b'}}\Jsym{-\sign(b)\abs{a'}\invint{b'}{N_b}}{N_{ab}}=
\nonumber \\
\lo= \Jsym{\abs{a'}}{N_{ab}\abs{b'}} \rme\left(-\sign(ab)
\left(N_{ab} + N_{ab}\abs{b'} - \abs{b'}- 1\right)/8\right).
\end{eqnarray*}

\subsection{Case 2: $N_a$ even and $a'$ odd}
\label{c2r} We shall follow the same technique as in
section~\ref{c1r}. The evaluation of
$\left[U_N(AS^+)\delta_0\right](0)$ is straightforward:
\begin{equation}
\label{mbet}
 \fl \left[U_N(AS^+)\delta_0\right](0) =
\sqrt{(a,N)}
\Jsym{\abs{a}}{\abs{b}}\Jsym{N_{ab}\abs{b'}}{\abs{a'}}
\rme\left(\sign(ab)\left(\abs{b} + \abs{a'} - 1 \right)/8\right).
\end{equation}
The computation of $\left[U_N(A)U_N(S^+)\delta_0\right](0)$ does
not differ from the previous case until equation~\eref{finsume},
which now becomes
\begin{eqnarray}
\label{finsume2}
 \fl \left[U_N(A)U_N(S^+)\delta_0\right](0)  =
 \sqrt{(a,N)}\Jsym{\abs{a}}{\abs{b}}\Jsym{N_{ab}\abs{a'}}{\abs{b'}}
 \Jsym{N_{ab}}{-\sign(b)\abs{a'}\invint{b'}{N_b}}\nonumber \\
 \times \rme\left(\sign (ab)\left(\abs{b}-\abs{b'}\right)/8
 + \invint{b'}{N_{b}}a'/8\right].
\end{eqnarray}
As in equation~\eref{finsume}, we have chosen $\invint{b'}{N_b}$
so that $\sign \left[\invint{b'}{N_b} \right]=- \sign(b)$.
Rearranging the Jacobi symbols gives
\begin{eqnarray*}
 \fl \Jsym{N_{ab}\abs{a'}}{\abs{b'}}
 \Jsym{N_{ab}}{-\sign(b)\abs{a'}\invint{b'}{N_b}} = \Jsym{N_{ab}\abs{b'}}{\abs{a'}}
\Jsym{N_{ab}}{-b'\invint{b'}{N_b}} \nonumber \\
\quad \times \rme\left(-\sign(ab) \left(\abs{a'}\abs{b'} -\abs{a'}
-\abs{b'} + 1\right)/8\right).
\end{eqnarray*}
Finally, the equality of the right-hand sides of
equations~\eref{mbet} and~\eref{finsume2} follows from
\[
 \Jsym{N_{ab}}{-b'\invint{b'}{N_b}}
\rme\left(a'\left[\invint{b'}{N_b}-b'\right]/8\right)=1,
\]
which can be obtained directly from property~\eref{mod-eight} of
the Jacobi symbol.

\subsection{Case 3: $N_a$ odd and $a'$ odd}
\label{c3r}

If we chose $\delta_0(Q)$ in this case too, the sum~\eref{finsum}
would be zero. Now the appropriate choice is
$\delta_{\frac{(a,N)}{2}}(Q)$. We have
\begin{eqnarray}
\label{simsum3} \fl
\left[U_N(AS^+)\delta_{\frac{(a,N)}{2}}\right](0)  = \sqrt{(a,N)}
\Jsym{\abs{a}}{\abs{b}}\Jsym{N_{ab}\abs{b'}}{\abs{a'}}\nonumber \\
\quad \times \rme\left(\sign(ab)\left(\abs{b} + \abs{a'} - 1
\right)/8 \right)
\rme\left(-\frac{b'}{8N_{ab}}\invint{a'}{8N_{ab}}\right).
\end{eqnarray}

On the other hand, the computation of the right-hand side of
equation~\eref{mult_t2} leads to
\begin{eqnarray}
\label{finsum3}
 \fl \left[U_N(A)U_N(S^+)\delta_{\frac{(a,N)}{2}} \right](0) =
 \sqrt{(a,N)}\Jsym{\abs{a}}{\abs{b}}
\Jsym{N_{ab}\abs{a'}}{\abs{b'}}
\Jsym{-\sign(b)\abs{a'}\invint{b'}{N_b}}{N_{ab}} \nonumber \\
\times \rme\left(\sign (ab) \left(\abs{b} - \abs{b'} + N_{ab} - 1
\right)/8\right)
 \rme\left(-\frac{1}{8N_{ab}}b'\invint{a'}{N_{ab}}4^{2\phi(N_{ab})}\right).
\end{eqnarray}
Using
\begin{eqnarray*}
\fl
\Jsym{N_{ab}\abs{a'}}{\abs{b'}}\Jsym{-\sign(b)\abs{a'}\invint{b'}{N_b}}{N_{ab}}
=\\
\lo=\Jsym{N_{ab}\abs{b'}}{\abs{a'}}\rme\left(\sign(ab)
\left(\abs{b'}+ \abs{a'}-N_{ab}\right)/8 - a'b'N_{ab}/8\right),
\end{eqnarray*}
equation~\eref{finsum3} simplifies to
\begin{eqnarray}
\label{llab}
 \fl \left[U_N(A)U_N(S^+)\delta_{\frac{(a,N)}{2}}\right](0)= \sqrt{(a,N)}
\; \rme\left(-\frac{1}{8N_{ab}} \;
b'\invint{a'}{N_{ab}}4^{2\phi(N_{ab})}\right)
 \nonumber \\
\times\Jsym{\abs{a}}{\abs{b}}\Jsym{N_{ab}\abs{b'}}{\abs{a'}}
\rme\left(\sign(ab)\left(\abs{b} + \abs{a'} - 1 \right)/8 - a' b'
N_{ab}/8 \right).
\end{eqnarray}
Since $(b',8N_{ab})=1$, the equality of the right-hand sides of
equations~\eref{simsum3} and~\eref{llab} follows from the
congruence
\[
 \invint{a'}{8N_{ab}} \equiv
\invint{a'}{N_{ab}}4^{2\phi(N_{ab})} + a'N_{ab}^2\,  \bmod
8N_{ab},
\]
which can be easily proved using the Chinese remainder theorem.

 The proof of theorem~\ref{th1} is now completed.

 \ack The author is grateful to David Farmer, Jon Keating, P\"ar
Kurlberg, Jens Marklof and Ze\'ev Rudnick for helpful discussions,
and to Brian Conrey, the director of the American Institute of
Mathematics, for kind hospitality while this work was completed
and written.

\appendix
\setcounter{section}{0}

\section{Some properties of the Jacobi symbol}

In this appendix we present the definitions and main properties of
the Legendre and Jacobi symbols. For a more complete exposition
the reader should consult a book on elementary Number Theory (see,
e.g.~\cite{IR90}).

Let $p$ be a prime greater than $2$ and $q \in \mathbb{Z}$.
 The {\it Legendre symbol} is defined by
\[
\Jsym{q}{p} := \cases{ 1 & if $\exists \, x \in
\mathbb{Z}/p\,\mathbb{Z}\,|$ $x^2 \equiv q \bmod p$,  \\
 -1 & if $\nexists \, x \in
\mathbb{Z}/p\, \mathbb{Z}\,|$ $x^2 \equiv q \bmod p$, \\
0 & if $p \mid q$.}
\]

The Legendre symbol has the following properties.
\begin{enumerate}
\item
\[
\Jsym{q_1q_2}{p} = \Jsym{q_1}{p}\Jsym{q_2}{p} \quad {\rm and}
\quad \Jsym{q}{p_1p_2}= \Jsym{q}{p_1}\Jsym{q}{p_2},
\]
\item
\[
\Jsym{1}{p} = 1 \quad {\rm and } \quad \Jsym{q + p}{p} =
\Jsym{q}{p},
\]
\item
\[
\Jsym{-1}{p} = \cases{1 & for $p \equiv 1 \bmod 4$ \\
                      -1 & for $p \equiv 3 \bmod 4$,}
\]
i.e.
\[
\Jsym{-1}{p} = \rme\left(\pm \left(p-1\right)/4\right),
\]
\item
\[
\Jsym{2}{p} = \cases{1 & for $p \equiv \pm 1 \bmod 8$ \\
                      -1 & for $p \equiv \pm 3 \bmod 8$,}
\]
which can be written
\[
\Jsym{2}{p} = \rme\left(\pm \left(p^2-1\right)/16\right),
\]
\item if $p$ and $q$ are both prime and greater than $2$, then
\[
\Jsym{p}{q}\Jsym{q}{p} = \cases{-1 & if $p \equiv q \equiv 3$ \\
                                 1 & otherwise,}
\]
or equivalently
\[
\Jsym{p}{q}\Jsym{q}{p}=\rme\left(\pm \left(p-1)(q - 1)
\right)/8\right).
\]
\end{enumerate}
Property (v) is known as the {\it quadratic reciprocity law}. The
sign in the arguments of the above exponentials is obviously
arbitrary. We make extensive use of this simple fact in this
paper.

Let $r$ be an odd positive integer and $q \in \mathbb{Z}$. Let $r
=p_1 p_2 \cdots p_n$ where the $p_i$ are (not necessarily
distinct) primes. The symbol
\[
\Jsym{q}{r} := \Jsym{q}{p_1}\Jsym{q}{p_2} \cdots \Jsym{q}{p_n}
\]
is called the {\it Jacobi symbol}.

Properties (i) and (ii) trivially extend to the Jacobi symbol.
Also (iii-v) apply to the Jacobi symbol (for reciprocity both
numbers must be odd and positive). This follows from the
congruences
\begin{eqnarray*}
\sum_{k=1}^n \frac{r_k - 1}{2} & \equiv \frac{r_1r_2 \cdots r_n-1}%
{2} \bmod 2, \\
\sum_{k=1}^n \frac{r_k^2 -1}{8} & \equiv
 \frac{ r_1^2r_2^2 \cdots r_n^2 -1}{8} \bmod 2,
\end{eqnarray*}
where $r_1, r_2,\ldots,r_n$ are odd integers.

Let $q$ and $r$ be positive odd integers and $r \equiv \pm 1 \bmod
4q$.  Using the reciprocity law and properties (ii) and (iii) one
easily obtains
\begin{equation}
\label{odd_one}
 \Jsym{q}{r} = 1.
\end{equation}
Now, let $q = 2^\alpha \bar{q}$, with $\alpha \ge 1$ and $\bar{q}$
odd, and let $r \equiv \pm 1 \bmod 4\bar{q}$. Then,
by~\eref{odd_one} and using (i) and (iv), we have
\begin{equation}
\label{mod-eight}
\Jsym{q}{r}= \cases{ -1 & if $q \equiv 2 \bmod
                                4$ and $r \equiv \pm 3 \bmod 8$, \\
                                1 & otherwise.}
\end{equation}

\section{Quantum mechanics on the two-torus}
\label{qmt}

We briefly review the quantum mechanics of systems whose classical
phase space is $\mathbb{T}^2$. For more details see, e.g.
\cite{HB80,Kna90,Esp93,DEG96,KR00}.  Without loss of generality,
we restrict to the case when $\boldsymbol{\varphi}=(0,0)$, i.e.
the wavefunction is exactly periodic in both the position and
momentum bases.

The periodicity of the wavefunction in both representations has
two important consequences. Firstly, both $\psi(q)$ and
$\hat\psi(p)$ are superpositions of delta functions supported on
the lattices points $q=2\pi\hbar \; Q$ and $p=2 \pi \hbar\; P$
respectively, where $Q,P \in \mathbb{Z}$. That is,
\[
\psi(q)=\sum_{m \in \mathbb{Z}} \sum_{Q=0}^{N-1} \Psi(Q) \,
\delta\left(q-\frac{Q}{N}+m\right)
\]
with $\Psi(Q +N)=\Psi(Q)$. Secondly, $2 \pi \hbar$ must be an
inverse integer, i.e. $N=1/2\pi\hbar$.  It follows that the
Hilbert space may be identified with the $N$-dimensional vector
space $\HN = \Ltwo(\mathbb{Z}/N\mathbb{Z})\cong \mathbb{C}^N$. For
$\nu = 0, \ldots, N - 1$ the set of functions
\begin{equation}
\label{b_f}
 \delta_\nu(Q) := \cases{\sqrt N  & if $Q \equiv \nu \bmod N$, \\
                        0  & if $Q\notequiv \nu \bmod N$}
\end{equation}
forms an orthogonal basis in $\HN$.

Quantization means mapping of observables $f\in
C^\infty(\mathbb{T}^2)$ into operators $\Op(f)$ acting on $\HN$
such that as $N \rightarrow \infty$, i.e. $\hbar \rightarrow 0$,
$\Op(f)$ tends to $f$.  More precisely, if $f,g \in
C^{\infty}(\mathbb{T}^2)$ then we require that
\begin{equation}
\label{cl_lim}
 \frac{N}{2\pi}\left[\Op(f),\Op(g)\right] -
\Op\left(\left \{f,g \right \} \right) \underset{N\rightarrow
\infty}{\longrightarrow} 0.
\end{equation}
Here $[ \cdot\, ,\cdot]$ denotes the commutator of two operators
and
\[
\left \{f,g\right \} = \frac{\partial f}{\partial p}\frac{\partial
g}{\partial q} -  \frac{\partial f}{\partial q}\frac{\partial
g}{\partial p}
\]
are the Poisson brackets of two observables.

In order to define a quantization satisfying~\eref{cl_lim}, we
need to introduce the translation operators
\begin{eqnarray*}
 t_1 \Phi(Q) & := \Phi(Q) \rme \left(\frac{Q}{N}\right),\\
 t_2 \Phi(Q)& := \Phi(Q+1),
\end{eqnarray*}
which may be thought of as the analogue of $\rme(\hat{q})$ and
$\rme(\hat{p})$, where $\hat{q}$ and $\hat{p}$ are the usual
position and momentum operators in $\Ltwo(\mathbb{R})$ . For any
$m,n \in \mathbb{Z}$, we have the following commutation relation
\[
t_1^m t_2^n= \rme \left(-\frac{mn}{N}\right)t_2^nt_1^m.
\]
Note that $t_1^N=t_2^N=1_N$.

The Weyl-Heisenberg operators are defined by
\[
T_N(\bi{n}):= \rme \left(-\frac{n_1n_2}{2N}\right)
t_2^{n_2}t_1^{n_1},
\]
where $\bi{n}=(n_1,n_2)$.  We then have the following
multiplication rule
\[
T_N(\bi{m})T_N(\bi{n})=
\rme\left(-\frac{1}{2N}\,\omega\left(\bi{m},\bi{n}\right)\right)
T_N(\bi{m} + \bi{n}),
\]
where $\omega(\bi{m},\bi{n}):= m_1n_2-m_2n_1$ is the standard
symplectic form.

Let $f \in C^{\infty}(\mathbb{T}^2)$ be a classical observable on
$\mathbb{T}^2$ whose Fourier series is given by
\[
f(\bi{z}) = \sum_{\bi{m} \in \mathbb{Z}^2} \hat f_{\bi{m}}\, \rme
\left(
  \bi{z}\cdot\bi{m}\right),
  \quad \bi{z}=\Biggl(\begin{array}{c} q \\p \end{array}\Biggl)
   \in \mathbb{T}^2.
\]
The Weyl quantization of $f$ is defined as
\[
 \Op(f) := \sum_{\bi{m} \in \mathbb{Z}^2}
\hat{f}_{\bi{m}} \, T_N(\bi{m}) \, .
\]
If $f$ is real, then $\Op(f)$ is self-adjoint. Moreover, it can be
shown that $\Op(f)$ satisfies~\eref{cl_lim}.

\section{Proof of equation~\eref{invp_h}}
In this appendix we show that
\[
h(a,b) = h(d,b), \quad A= \Biggl(\begin{array}{cc}
      a & b \\ c & d
     \end{array} \Biggr) \in \SLTZ,
\]
where the function $h$ is defined in~\eref{hf}.
 If $a$ is even we have
\[
h(a,b) =
\Jsym{\sign(ad)}{\abs{b}}\Jsym{\abs{d}}{\abs{b}}\rme\left(\sign(ab)\left(\abs{b}
- 1\right)/8\right).
\]
If $\sign(ad)=1$ then $\sign(ab)=\sign(bd)$ and~\eref{invp_h}
follows immediately.  If $\sign(ad)=-1$ then
\[
\fl \Jsym{-1}{\abs{b}}\rme\left(\sign(ab)\left(\abs{b} -
1\right)/8\right) =\rme\left(-\sign(ab)\left(\abs{b} -
1\right)/8\right) =\rme\left(\sign(bd)\left(\abs{b} -
1\right)/8\right),
\]
which proves~\eref{invp_h}.

Now, let  $a$ be an odd number and $b=2^\alpha\bar{b}$, where
$\alpha \ge 1$ and $\bar{b}$ is odd.  The condition $ad-bc=1$
implies that $\abs{ad} \equiv \pm 1 \bmod 4\abs{\bar{b}}$.
Moreover, we have either
\begin{equation}
\label{latc1}
 a \equiv d \bmod 8
\end{equation}
or
\begin{equation}
\label{latc}
 d \equiv a + 4 \bmod 8.
\end{equation}
The congruence~\eref{latc1} implies
\[
h(a,b)=\Jsym{\abs{b}}{\abs{ad}}\Jsym{\abs{b}}{\abs{d}}
\rme\left(-\sign(ab)\abs{a}/8\right) = h(d,b),
\]
which follows from property~\eref{mod-eight} of the Jacobi symbol
and from the equivalence $ad \equiv 1 \bmod 8$.  If~\eref{latc}
holds, then $ad \equiv -3 \bmod 8$ and $b = 2\bar{b}$. Therefore,
we have
\begin{equation}
\label{eqa}
 \Jsym{\abs{b}}{\abs{ad}} =
\Jsym{2}{\abs{ad}}\Jsym{\abs{\bar{b}}}{\abs{ad}} = - 1.
\end{equation}
Combining~\eref{latc} and~\eref{eqa} we obtain
\[
h(a,b) = \Jsym{\abs{b}}{\abs{d}}\rme\left(-\sign(ab)\left(\abs{a}
+ 4\right)/8\right) = h(d,b).
\]

\end{document}